\documentclass [] {aa} 
\usepackage{epsfig, graphicx,longtable}
\usepackage{natbib}
\usepackage{times}

\bibpunct{(}{)}{;}{a}{}{,}

\begin{document}


\title{Statistical properties of exoplanets\thanks{Based on 
               observations collected at the La Silla Observatory, 
               ESO (Chile), with the CORALIE spectrograph 
               at the 1.2-m Euler Swiss telescope and the FEROS spectrograph
               at the 1.52-m ESO telescope, with the VLT/UT2 
               Kueyen telescope (Paranal Observatory, ESO, Chile) using the 
               UVES spectrograph (Observing run 67.C-0206, in service 
               mode), with the 
               TNG and William Herschel Telescopes, both operated at the 
               island of La Palma, and with the ELODIE spectrograph at the
               1.93-m telescope at the Observatoire de Haute Provence.}}
\subtitle{II. Metallicity, orbital parameters, and space velocities}

\author{
      N.~C.~Santos \inst{1}
 \and G.~Israelian \inst{2}
 \and M.~Mayor \inst{1}
 \and R.~Rebolo \inst{2,3}
 \and S.~Udry \inst{1}
}

\offprints{Nuno C. Santos, \email{Nuno.Santos@obs.unige.ch}}

\institute{
	Observatoire de Gen\`eve, 51 ch.  des 
	Maillettes, CH--1290 Sauverny, Switzerland
     \and
	Instituto de Astrof{\'\i}sica de Canarias, E-38200 
        La Laguna, Tenerife, Spain
     \and
        Consejo Superior de Investigaciones Cient\'{\i}ficas, Spain}
\date{Received 19-09-2002/ Accepted 07-11-2002 -- IN PRESS -- IN PRESS -- IN PRESS -- IN PRESS} 

\titlerunning{} 


\abstract{
In this article we present a detailed spectroscopic analysis of more than 50 extra-solar planet
host stars. Stellar atmospheric parameters and metallicities are derived using high
resolution and high S/N spectra. The spectroscopy results, added to the previous studies,
imply that we have access to a large and uniform sample of metallicities for about 80 planet hosts
stars. We make use of this sample to confirm the metal-rich nature of stars with planets,
and to show that the planetary frequency is rising as a function of the [Fe/H].
Furthermore, the source of this high metallicity is shown to have most probably an
``primordial'' source, confirming previous results. The comparison of the orbital properties 
(period and eccentricity) and minimum masses of the planets with the stellar properties 
also reveal some emerging but still not significant trends. These are discussed and some 
explanations are proposed. Finally, we show that the planet host stars included in the CORALIE 
survey have similar kinematical properties as the whole CORALIE volume-limited planet search 
sample. Planet hosts simply seem to occupy the metal-rich envelope of this latter population.
}

\maketitle

\section{Introduction}

The discovery of now more than 100 extra-solar giant 
planets\footnote{See e.g. tables at http://obswww.unige.ch/exoplanets/} 
opened a wide range of questions regarding the understanding of the mechanisms 
of planetary formation. To find a solution for the many problems risen, we need 
observational constraints. These can come, for example, from the analysis of 
orbital parameters of the known planets, like the distribution of planetary 
masses \citep[][]{Jor01,Zuc01,Udr01}, eccentricities, or orbital periods \citep[for 
a review see e.g.][]{Udr01,MaySan02}.
In fact, as new and longer period planets are found, more interesting correlations are 
popping up.  Examples of these are the discovery that there is a paucity of high mass companions
orbiting in short period orbits \citep[][]{Zuc02,Udr02}, or
the interesting lack of long-period low-mass planets \citep[][]{Udr02b}. 

But further evidences are coming from the study of the planet host stars themselves.
Precise spectroscopic studies have revealed that stars with planets seem to be 
particularly metal-rich when compared with ``single'' field dwarfs
\citep[][]{Gon97,Fur97,Gon98,San00,Gon01,San01a,San01b}. 
Furthermore, the frequency of planets seems to be a strong function of 
[Fe/H] \citep[][ hereafter Paper\,II]{San01a}. These facts, that were shown not to result 
from any sampling bias (Paper II), 
are most probably telling us that the metallicity plays a key role in the formation of
a giant planet, or at least of a giant planet like the ones we are finding now.

The source of this metallicity ``excess'' has, however, been a matter of debate.
Some authors have suggested that the high metal content of the planet host stars
may have an external origin: it results from the addition of metal-rich (hydrogen poor)
material into the convective envelope of the star, a process that could result from
the planetary formation process itself \citep[][]{Gon98, Lau00, Gon01, Smi01, Mur02}.
Evidences for the infall of planetary material have in fact been found for
a few planet host stars \citep[e.g.][]{Isr01,Law01,Isr02}, although not necessarily
able to change considerably the overall metal content \citep[see e.g.][]{Sand02}.
In fact, most evidences today suggest that the metallicity ``excess'' as a whole
has a ``primordial'' origin \citep[][]{Pin01,San01a,San01b,San02,Sad02},
and thus that the metal content of the cloud giving birth to the
star and planetary system is indeed a key parameter to form a giant planet.

Besides the simple correlation between the presence of a planet and the
high metal-content of its host star, there are some hints that the
metallicity might be correlated with the planetary orbital properties. 
For example, \citet[][]{Gon98} and \citet[][]{Que00} have shown
some evidences that stars with very short-period planets (i.e. small semi-major axes) 
may be particularly metal-rich, even amongst the planetary hosts. More recent studies
have, however, failed to confirm this relation (Paper~II).

With the number of new planets growing every day,
it is extremely important to survey the samples for new emerging correlations
between the stellar properties and the characteristics (minimum masses and orbital
parameters) of the planetary
companions. In this paper we focus exactly on this point. We have obtained high-resolution 
and high-S/N spectra of more than 50 extra-solar planet host stars, most of them without 
any previous detailed spectroscopic analysis. These new determinations bring to about 
80 our sample of planet host stars with homogeneous derived spectroscopic parameters.
The structure of this article goes as follows. In Sect.\,\ref{sec:data} we 
describe the observations and the chemical analysis. In Sect.\,\ref{sec:metallicity} we 
review the current status of the metallicity distribution of stars with planets, 
further discussing its origin, and in Sect.\,\ref{sec:parameters} we analyze the relation 
between the orbital parameters and the metallicity. In Sect.\,\ref{sec:uvw} we finally 
analyze the space velocities of planet and non-planet host stars, comparing them with 
the stellar metallicity. We conclude in Sect.\,\ref{sec:conclusions}.

\section{Observations and spectroscopic analysis}
\label{sec:data}

\subsection{Observations and data reduction}

High resolution spectra for more than 50 planet hosts stars were obtained during several
runs using 6 different spectrographs. The general characteristics of these instruments
are presented in Table\,\ref{tabspec}.

\begin{table}
\caption[]{Spectrographs used for the current study and their spectral coverage and resolution}
\begin{tabular}{lcc}
\hline
Spectrograph/Telescope               & Resolution                  & Coverage     \\
                                     & ($\lambda$/$\Delta\lambda$) & (\AA)        \\
\hline
CORALIE/1.2-m Euler Swiss   & 50\,000                     & 3\,800-6\,800    \\
FEROS/1.52-m ESO            & 48\,000 		  	  & 3\,600-9\,200    \\
UES/4-m William Hershel     & 55\,000 			  & 4\,600-7\,800    \\
SARG/3.5-m TNG              & 57\,000 			  & 5\,100-10\,100   \\
UVES/VLT 8-m Kueyen UT2     & 110\,000 			  & 4\,800-6\,800    \\
ELODIE/1.93-m OHP           & 48\,000                     & 3\,800-6\,800    \\
\hline
\end{tabular}
\label{tabspec}
\end{table}

The spectra have in general a S/N ratio between 150 and 400, but are as high as
1\,000 for the UVES spectra. Except for the UES spectra, all the others cover very well
all the spectral domain without any significant gaps, permitting us to measure the 
Equivalent Widths for most of the spectral lines used 
(see \citet[][]{San00} -- hereafter Paper\,I -- and Paper\,II). But even in this case, the gaps did not imply any strong limitations, since the available lines still 
have a wide variety of equivalent widths and lower excitation potentials, essential to
the precise determination of the stellar parameters (see next section).

\begin{table*}
\caption[]{Stellar parameters derived in the current study. $\xi_t$ denotes the microturbulence
parameters. For a list of the planet 
discovery papers see tables at http://obswww.unige.ch/exoplanets/
and http://cfa-www.harvard.edu/planets/. A more complete table with the number 
of \ion{Fe}{i} and \ion{Fe}{ii} lines used and the dispersions will be available at CDS}
\begin{tabular}{lcccrlc|lcccrlc}
\hline
Star         & T$_{\mathrm{eff}}$ & $\log{g}$ & $\xi_t$      & [Fe/H] & Inst. & M$_\star^{\dag\dag\dag}$ & Star  & T$_{\mathrm{eff}}$ & $\log{g}$ & $\xi_t$      & [Fe/H] & Inst. & M$_\star^{\dag\dag\dag}$\\
             &              (K) &     (cgs)&  {\small (km\,s$^{-1}$)} &        &  ($^a$)      & (M$_{\sun}$)                &       &              (K) &     (cgs)&  {\small (km\,s$^{-1}$)} &        &    ($^a$)       & (M$_{\sun}$)        \\
\hline
\object{\footnotesize{HD}\,142   } &6290 &4.38 &1.91 &0.11 &[2] &1.26                & \object{\footnotesize{HD}\,117176} &5530 &4.05 &1.08 &$-$0.05 &[4] &0.92\\
\object{\footnotesize{HD}\,2039  } &5990 &4.56 &1.24 &0.34 &[1] &1.20                & \object{\footnotesize{HD}\,128311} &4950 &4.80 &1.00 &0.10 &[3] &0.76\\
\object{\footnotesize{HD}\,4203  } &5650 &4.38 &1.15 &0.40 &[2] &0.93                & \object{\footnotesize{HD}\,130322} &5430 &4.62 &0.92 &0.06 &[4] &1.04\\
\object{\footnotesize{HD}\,4208  } &5625 &4.54 &0.95 &$-$0.23 &[2] &0.86             & \object{\footnotesize{HD}\,134987} &5780 &4.45 &1.06 &0.32 &[4] &1.05\\
\object{\footnotesize{HD}\,8574  } &6080 &4.41 &1.25 &0.05 &[4] &1.17                & \object{\footnotesize{HD}\,136118} &6175 &4.18 &1.61 &$-$0.06 &[4] &1.28\\
\object{\footnotesize{HD}\,9826  } &6120 &4.07 &1.50 &0.10 &[4] &1.29                & \object{\footnotesize{HD}\,137759} &4750 &3.15 &1.78 &0.09 & [4] & --\\ 
\object{\footnotesize{HD}\,10697 } &5665 &4.18 &1.19 &0.14 &[4] &1.22                & \object{\footnotesize{HD}\,141937} &5925 &4.62 &1.16 &0.11 &[3]$^{\dag\dag}$ &1.10\\
\object{\footnotesize{HD}\,12661 } &5715 &4.49 &1.09 &0.36 &[3] &1.05                & \object{\footnotesize{HD}\,143761} &5835 &4.40 &1.29 &$-$0.21 &[4] &0.95\\
\object{\footnotesize{HD}\,19994 } &6165 &4.13 &1.49 &0.23 & [1]$^\dag$ &1.34	     & \object{\footnotesize{HD}\,145675} &5255 &4.40 &0.68 &0.51 &[4] &0.90\\
                                   &6250 &4.27 &1.56 &0.30 & [2]$^\dag$ &1.35        & \object{\footnotesize{HD}\,147513} &5880 &4.58 &1.17 &0.07 &[1] &1.11\\
                                   &6105 &4.02 &1.51 &0.18 & [5] &1.34               & \object{\footnotesize{HD}\,150706} &6000 &4.62 &1.16 &0.01 &[3] &1.21\\
     (average)                     &6175 &4.14 &1.52 &0.21 & &1.34                   & \object{\footnotesize{HD}\,160691} &5820 &4.44 &1.23 &0.33 &[1]$^{\dag\dag}$ &1.10\\
\object{\footnotesize{HD}\,20367 } &6100 &4.55 &1.31 &0.14 & [6] &1.17               & \object{\footnotesize{HD}\,168443} &5600 &4.30 &1.18 &0.06 &[4] &0.96\\
\object{\footnotesize{HD}\,23079 } &5945 &4.44 &1.21 &$-$0.11 &[2] &1.00             & \object{\footnotesize{HD}\,177830} &4840 &3.60 &1.18 &0.32 &[4] &1.03\\
\object{\footnotesize{HD}\,23596 } &6125 &4.29 &1.32 &0.32 &[3] &1.30                & \object{\footnotesize{HD}\,179949} &6235 &4.41 &1.38 &0.21 &[1]$^{\dag\dag}$ &1.25\\
\object{\footnotesize{HD}\,27442 } &4890 &3.89 &1.24 &0.42 &[2] &0.83                & \object{\footnotesize{HD}\,186427} &5765 &4.46 &1.03 &0.09 &[4] &0.99\\
\object{\footnotesize{HD}\,30177 } &5590 &4.45 &1.07 &0.39 &[1] &1.00                & \object{\footnotesize{HD}\,187123} &5855 &4.48 &1.10 &0.14 &[4] &1.05\\
\object{HD\,33636 } &5990 &4.68 &1.22 &$-$0.05 &[2] &1.12                            & \object{\footnotesize{HD}\,190228} &5360 &4.02 &1.12 &$-$0.24 &[3]$^\dag$ &0.84\\
\object{HD\,37124 } &5565 &4.62 &0.90 &$-$0.37 &[3] &0.76                            &  				  &5325 &3.95 &1.10 &$-$0.23 &[4] &0.82\\
\object{\footnotesize{HD}\,39091 } &5995 &4.48 &1.30 &0.09 &[1]$\dag$ &1.10          &   (average)   &5340 &3.99 &1.11 &$-$0.24 &  &0.83\\
\object{\footnotesize{HD}\,46375 } &5315 &4.54 &1.11 &0.21 &[3] &0.83                & \object{\footnotesize{HD}\,190360} &5590 &4.48 &1.06 &0.25 &[3] &0.96\\
\object{\footnotesize{HD}\,50554 } &6050 &4.59 &1.19 &0.02 &[3] &1.11                & \object{\footnotesize{HD}\,192263}$^{\star}$ &4995 &4.76 &0.90 &0.04 &[2] &0.75\\
\object{\footnotesize{HD}\,74156 } &6105 &4.40 &1.36 &0.15 &[2] &1.27                & \object{\footnotesize{HD}\,195019} &5845 &4.39 &1.23 &0.08 &[4] &1.06\\
\object{\footnotesize{HD}\,75732A} &5307 &4.58 &1.06 &0.35 &[3] &0.88                &  	     &5832 &4.34 &1.24 &0.09 &[1]$^{\dag\dag}$ &1.05\\
\object{\footnotesize{HD}\,80606 } &5570 &4.56 &1.11 &0.34 &[3] &1.03                &   (average)   &5840 &4.36 &1.24 &0.08 &   &1.06\\
\object{\footnotesize{HD}\,82943 } &6025 &4.54 &1.10 &0.33 &[1]$^\dag$ &1.15         & \object{\footnotesize{HD}\,196050} &5905 &4.41 &1.40 &0.21 &[1] &1.10\\
                    &6025 &4.53 &1.15 &0.30 &[5] &1.15                               & \object{\footnotesize{HD}\,209458} &6120 &4.56 &1.37 &0.02 &[5] &1.15\\
       (average)    &6025 &4.54 &1.12 &0.32 &    &1.15                               & \object{\footnotesize{HD}\,210277} &5575 &4.44 &1.12 &0.23 &[2]$^\dag$ &0.94\\
\object{\footnotesize{HD}\,92788 } &5820 &4.60 &1.12 &0.34 &[1] &1.10                &  	     &5560 &4.46 &1.03 &0.21 &[4] &0.93\\
\object{\footnotesize{HD}\,95128 } &5925 &4.45 &1.24 &0.05 &[4] &1.05                &   (average)   &5570 &4.45 &1.08 &0.22 &  &0.94\\
\object{\footnotesize{HD}\,106252} &5890 &4.40 &1.06 &$-$0.01 &[1]$^{\dag\dag}$ &1.02& \object{\footnotesize{HD}\,213240} &5975 &4.32 &1.30 &0.16 &[1] &1.22\\
\object{\footnotesize{HD}\,108874} &5615 &4.58 &0.93 &0.25 &[3] &0.96                & \object{\footnotesize{HD}\,216435} &5905 &4.16 &1.26 &0.22 &[1] &1.26 \\
\object{\footnotesize{HD}\,114386} &4875 &4.69 &0.63 &0.00 &[1] &0.68                & \object{\footnotesize{HD}\,216437} &5875 &4.38 &1.30 &0.25 &[1] &1.06\\
\object{\footnotesize{HD}\,114729} &5820 &4.20 &1.03 &$-$0.26 &[3] &0.94             & \object{\footnotesize{HD}\,217014} &5805 &4.51 &1.22 &0.21 &[2] &1.04\\
\object{\footnotesize{HD}\,114762} &5870 &4.25 &1.28 &$-$0.72 &[5] &0.80             & \object{\footnotesize{HD}\,222582} &5850 &4.58 &1.06 &0.06 &[3] &1.02\\
\object{\footnotesize{HD}\,114783} &5160 &4.75 &0.79 &0.16 &[4] &0.88                & & & & & & & \\\hline

\end{tabular}
\newline
{$^a$ The instruments are [1] CORALIE, [2] FEROS, [3] UES, [4] SARG, [5] UVES, and [6] ELODIE
\newline
$\dag$ Already published in \citet[][]{San01a} (Paper\,II)
\newline
$\dag\dag$ Already published in \citet[][]{San01b}
\newline
$\dag\dag\dag$ From the isochrones of \citet[][]{Sch92}, \citet[][]{Sch93} and \citet[][]{Schae92}
\newline
$\star$ The existence of a planet around HD192263 was recently put in cause by \citet[][]{Hen02};
we think, however, that these authors have not shown enough evidences against the 
presence of a planet, and thus we prefer to keep this star in the planet-hosts sample 
(Santos et al., in preparation)}
\label{tab1}
\end{table*}

Data reduction was done using IRAF\footnote{IRAF 
is distributed by National Optical Astronomy Observatories, operated 
by the Association of Universities for Research in Astronomy, Inc., 
under contract with the National Science Foundation, U.S.A.} tools 
in the {\tt echelle} package. Standard background correction, 
flat-field, and extraction procedures were used. In all the cases, the 
wavelength calibration was done using a ThAr lamp spectrum taken during the 
same night. 

We have compared the Equivalent Widths (EW) for some stars for which we have obtained spectra
using different instruments to check for possible systematics. In all cases,
the average difference of the EWs is within 1-2\,m\AA, and usually lower than 1\,m\AA.
As can be also verified from Table\,\ref{tab1} of this article and Table\,2 from Paper\,II, 
these possible small systematics does not seem to affect significantly the analysis of the
atmospheric parameters and metallicity. The only star that has large variations in the
derived atmospheric parameters is \object{HD\,19994}. This variation
might be connected with the fact that this late F dwarf has a high rotational 
velocity $v\,\sin{i}$=8.1\,km\,s$^{-1}$ 
\citep[from the calibration of the CORALIE Cross-Correlation Function presented in][]{San02a}, 
a sign of relative youth and (most probably) activity related phenomena.

\subsection{Stellar parameters and chemical analysis}

In this paper we use the same technique, line-lists, and model atmospheres
as in Papers~I and II. The abundance analysis was done in standard Local 
Thermodynamic Equilibrium (LTE) using a revised version of the 
code MOOG \citep{Sne73}, and a grid of \citet{Kur93} ATLAS9 atmospheres.

The atmospheric parameters were obtained from the \ion{Fe}{i} and \ion{Fe}{ii} lines
by iterating until the correlation coefficients between $\log{\epsilon}$(\ion{Fe}{i}) 
and $\chi_l$, and between $\log{\epsilon}$(\ion{Fe}{i}) and  $\log{({W}_\lambda/\lambda)}$ 
were zero, and the mean abundance given by \ion{Fe}{i} and \ion{Fe}{ii} lines were the same. 
This procedure gives very good results since the set of \ion{Fe}{i} lines has a very wide range 
of excitation potentials.

The results of our analysis are presented in Table\,\ref{tab1}. The number 
of measured \ion{Fe}{i} and \ion{Fe}{ii} lines is always between 24 and 39, and 4 and 8, respectively.
The rms around the mean individual abundances given by the lines has values between 0.03 and
0.07\,dex in most cases.
The errors in $T_\mathrm{eff}$, $\log{g}$, $\xi_t$ and [Fe/H] were computed as in \citet[][]{GonVan98}. For 
a typical measure the uncertainties are usually lower than 50\,K, 0.15\,dex, 0.10\,km\,s$^{-1}$, 
and 0.06\,dex, respectively (see Paper\,II)\footnote{These represent relative errors, and not absolute ones.}. 
The only important exceptions are the cases of \object{HD\,20367} (for which the 
lower quality ELODIE spectra with S/N$\sim$80-100 were responsible for errors of the order of 100\,K, 0.20\,dex, 0.15\,km\,s$^{-1}$, and 0.10\,dex in $T_\mathrm{eff}$, $\log{g}$, $\xi_t$ and [Fe/H], respectively) and for \object{HD\,137759}, a giant star for which the dispersion in the [Fe/H] values for individual lines was quite 
high\footnote{In this case the estimated errors in $T_\mathrm{eff}$, $\log{g}$, $\xi_t$ and 
[Fe/H] are of 150\,K, 0.40\,dex, 0.15\,km\,s$^{-1}$, and 0.17\,dex, respectively.}.
The masses were then determined from the theoretical isochrones of \citet{Sch92}, 
\citet[][]{Sch93} and \citet{Schae92}, using $M_{V}$ computed from Hipparcos 
parallaxes \citep[][]{ESA97} and  $T_{\mathrm{eff}}$ obtained from spectroscopy. 
We adopt a typical error of 0.1\,M$_{\sun}$ for the masses.

\begin{figure*}[t]
\psfig{width=17cm,file=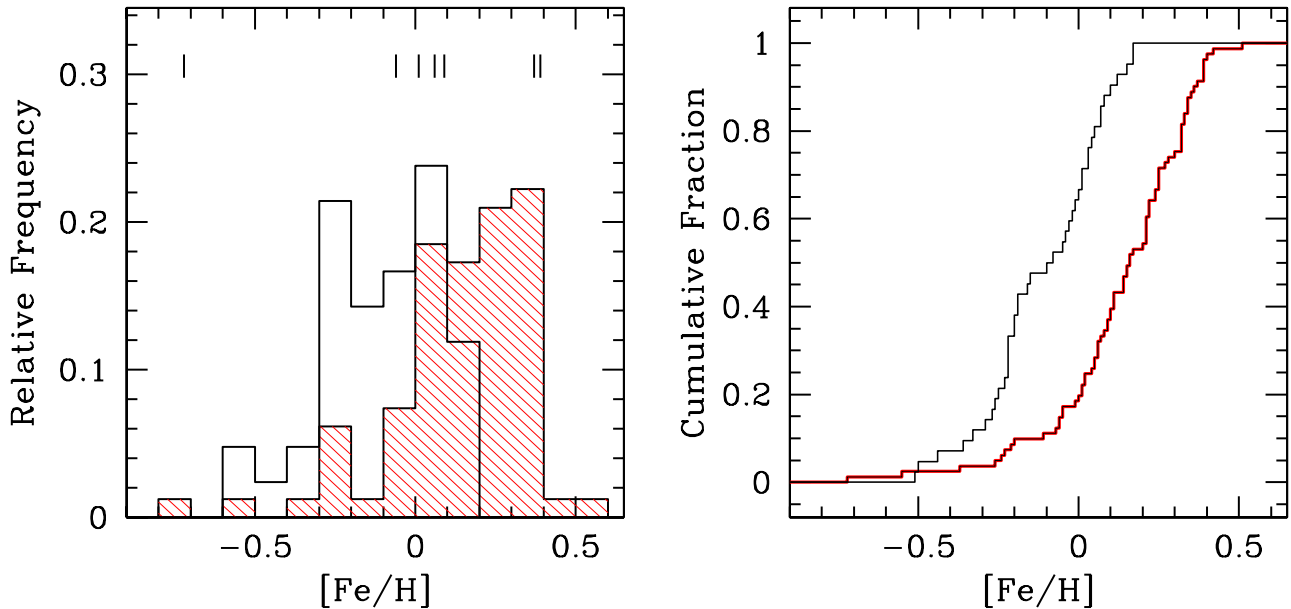}
\caption[]{{\it Left}: metallicity distribution for stars with planets (hashed histogram) compared 
with the same distribution for the field dwarfs presented in Paper\,II (empty histogram). 
The vertical lines represent stars with brown dwarf candidate companions. {\it Right}: The 
cumulative functions of both samples. A Kolmogorov-Smirnov test shows 
the probability for the two populations being part of the same sample is around 10$^{-7}$.}
\label{fig1}
\end{figure*}

\section{Confirming the excess metallicity}
\label{sec:metallicity}

In Fig.\,\ref{fig1} we plot the metallicity distribution for all the stars known to have companions with 
minimum masses lower than $\sim$18\,M$_{\mathrm{Jup}}$ (hashed histogram) 
when compared to the same distribution for a volume limited sample of stars with no (known) planetary 
companions (open histogram) -- see Paper\,II. 

For the planet host stars, most of the metallicity values ([Fe/H]) were taken from Table\,\ref{tab1}\footnote{The only 
star not used was \object{HD137759} (a K giant); we prefer to keep it out in the rest of the analysis.}
and from Table\,2 of Paper\,II. \object{HD\,39091}, a star that was included in the
``single'' star comparison sample in Paper\,II, was recently discovered to harbor
a brown dwarf companion \citep[][]{Jon02}. This star was thus taken out from this latter sample, 
and included in the planet sample. For 4 other stars for which we could not obtain spectra, values 
that were computed using the same technique (and are thus compatible with ours) are
available. These are for \object{BD\,$-$10\,3166} \citep[][]{Gon01}, \object{HD\,89744} \citep[][]{Gon01}, 
\object{HD\,120136} \citep[][]{Gon00}, and \object{HD\,178911\,B} \citep[][]{Zuc02b}. The parameters
(T$_{\mathrm{eff}}$, $\log{g}$, $\xi_t$, [Fe/H]) listed by these authors are (5320, 4.38, 0.85, 0.33),
(6338, 4.17, 1.55, 0.30), (6420, 4.18, 1.25, 0.32), and (5650, 4.65, 0.85, 0.28), respectively. 

We note that from the 89 stars known to harbor low mass (planetary or brown dwarf) 
companions, only 7 lack a metallicity determination\footnote{These are: 
\object{GJ\,876} (an M dwarf), \object{HD\,40979}, 
\object{HD\,49674}, \object{HD\,68988}, \object{HD\,72659}, \object{HD\,73526},
and \object{HD\,76700}.}.

It is important to remember that the metallicity for the two samples of stars plotted in Fig.\,\ref{fig1} 
was derived using exactly the same method, and are thus both in the same scale (Paper\,II).
This plot thus clearly confirms the already known trend that stars with planetary companions
are more metal-rich (in average) that field dwarfs. The average metallicity difference between the two 
samples is about 0.24\,dex, and the probability that the two distributions belong to the same sample
is of the order of 10$^{-7}$.

In Fig.\,\ref{fig1}, stars having ``planetary'' companions with masses higher than 10\,M$_{\mathrm{Jup}}$
are denoted by the vertical lines. Given the still low number, it is impossible to do any 
statistical study of this group. It is interesting though to mention that two of the
stars having companions in this mass regime (\object{HD\,202206} and \object{HD\,38529}) 
have very high metallicities ([Fe/H]=0.39 and 0.37, respectively), while one other (\object{HD\,114762})
has the lowest metallicity of all the objects in the sample ([Fe/H]=$-$0.72). This large dispersion might 
be seen as an evidence that (at least part of) the higher mass objects were formed by the same physical
mechanisms as their lower mass counterparts; the 13\,M$_{\mathrm{Jup}}$ deuterium burning limit has no
role in this matter. Furthermore,
this represents a good example to show that the metallicity by itself cannot be used 
as a planetary identification argument,
contrarily to what has been used by \citet[][]{Che01}.

We have tried to see if there were any differences between those stars
having multiple planetary systems and stars having one single planet. The analysis revealed no
significant trend. This negative result is probably not only due to the low number of points
used, but most of all to the fact that as can be seen in the literature \citep[e.g.][]{Fis01}, 
many planet host stars do present long term trends, some of them that might be induced by other planetary 
companions.

\begin{figure*}[t]
\psfig{width=17cm,file=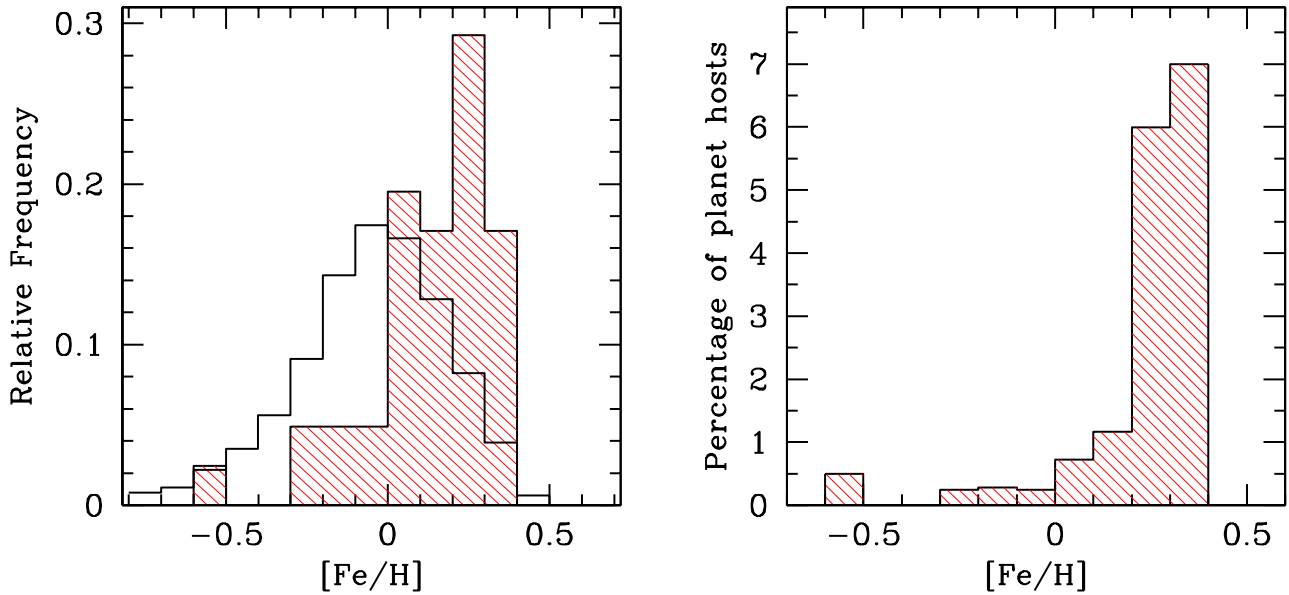}
\caption[]{{\it Left}: metallicity distribution of stars with planets making part of the
CORALIE planet search sample (shaded histogram) compared 
with the same distribution for the about 1000 non binary stars in the CORALIE volume-limited 
sample (see text for more details). {\it Right}: the percentage of stars belonging to the 
CORALIE search sample that have been discovered to harbor planetary mass companions 
plotted as a function of the metallicity. The vertical axis represents the percentage of
planet hosts with respect to the total CORALIE sample.}
\label{fig2}
\end{figure*}

\subsection{The probability of planet formation}
\label{sec:probability}

More interesting conclusions can be taken by looking at the shape of the distribution
of stars with planets. As it as been discussed in Paper\,II, 
this distribution is rising with [Fe/H], up to a value of $\sim$0.4, 
after which we see a sharp cutoff. This cutoff suggests that we may be looking at the approximate 
limit on the metallicity of the stars in the solar neighborhood. 

Here we have repeated the analysis presented in Paper\,II, but using only the
planet host stars included in the well defined CORALIE sample\footnote{These are: 
\object{HD\,142}, \object{HD\,1237}, \object{HD\,4208}, \object{HD\,6434}, \object{HD\,13445},
\object{HD\,16141}, \object{HD\,17051}, \object{HD\,19994}, \object{HD\,22049}, \object{HD\,23079},
\object{HD\,28185}, \object{HD\,39091}, \object{HD\,52265}, \object{HD\,75289}, \object{HD\,82943},
\object{HD\,83443}, \object{HD\,92788}, \object{HD\,108147}, \object{HD\,114386}, \object{HD\,114729},
\object{HD\,114783}, \object{HD\,121504}, \object{HD\,130322}, \object{HD\,134987}, \object{HD\,141937},
\object{HD\,147513}, \object{HD\,160691}, \object{HD\,162020}, \object{HD\,168443}, \object{HD\,168746},
\object{HD\,169830}, \object{HD\,179949}, \object{HD\,192263}, \object{HD\,196050}, \object{HD\,202206},
\object{HD\,210277}, \object{HD\,213240}, \object{HD\,216435}, \object{HD\,216437}, \object{HD\,217107}, and
\object{HD\,222582}.}. This sub-sample has a total of 41 objects, $\sim$60\% of them having planets 
discovered in the context of the CORALIE survey itself. Here we have included all stars
known to have companions with minimum masses lower than $\sim$18\,M$_{\mathrm{Jup}}$; changing this limit to e.g.
10\,M$_{\mathrm{Jup}}$ does not change any of the results presented below. 

The fact that planets seem to orbit the most metal-rich stars in the solar neighborhood has led some groups to
build planet search samples based on the high metal content of their host stars.
Examples of these are the stars \object{BD-10\,3166} \citep{But00}, \object{HD\,4203} \citep[][]{Vog02}, and
\object{HD\,73526}, \object{HD\,76700}, \object{HD\,30177}, and \object{HD\,2039} \citep[][]{Tin02}. 
Although clearly increasing the planet detection rate, these kind of metallicity biased 
samples completely spoil any statistical study. Using only stars being surveyed
for planets in the context of the CORALIE survey (none of these 6 stars is included), a survey 
that has never used the metallicity as a favoring quantity for looking for planets, has thus the
advantage of minimizing this bias.

As we can see from Fig.\,\ref{fig2} (left panel), the metallicity distribution for the planet host stars included
in the CORALIE sample does show an increasing trend with [Fe/H]. In the figure, the empty histogram 
represents the [Fe/H] distribution for a large volume limited sample of stars included in the CORALIE 
survey \citep[][]{Udr00}. The metallicities for this latter sample were computed from a precise calibration 
of the CORALIE Cross-Correlation Function \citep[see][]{San02a}; since the calibrators used were the stars 
presented in Paper\,I, Paper\,II, and this paper, the final results are in the very same scale.

The knowledge of the metallicity distribution for stars in the solar neighborhood 
(and included in the CORALIE sample) permits us to determine the percentage of planet 
host stars per metallicity bin. The result is seen in Fig.\,\ref{fig2} (right panel). 
As we can perfectly see, the probability of finding a 
planet host is a strong function of its metallicity. This result confirms former analysis done in 
Paper\,II and by \citet[][]{Rei02}. For example, here we can see that at about 7\% of the stars 
in the CORALIE sample having metallicity between 0.3 and 0.4\,dex have been discovered to harbor 
a planet. On the other hand, less than 1\% of the stars having solar metallicity seem to 
have a planet. This result is thus probably telling us that the probability of forming a 
giant planet, or at least a planet of the kind we are finding now, depends strongly on the
metallicity of the gas that gave origin to the star and planetary system. This might be 
simple explained if we consider that the higher the metallicity (i.e. dust density of the disk) 
the higher might be the probability of forming a core (and an higher mass core) before the 
disk dissipates \citep[][]{Pol96,Kok02}.

\begin{figure}[t]
\psfig{width=\hsize,file=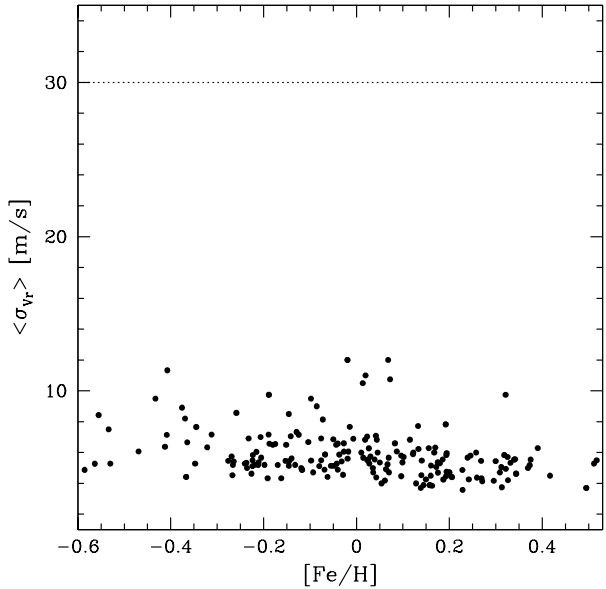}
\caption[]{Plot of the mean-photon noise error for the CORALIE measurements of stars 
having magnitude V between 6 and 7,
as a function of the metallicity. This latter quantity was computed using the calibration presented in \citet[][]{San02a}. Only a very few planet host stars present radial-velocity variations with
an amplitude smaller than 30\,m\,s$^{-1}$ (dotted line).}
\label{fig_errors}
\end{figure}

Although it is unwise to draw any strong conclusions based on only one point, it is worth noticing that
our own Sun is in the ``metal-poor'' tail of the planet host [Fe/H] distribution. 
Other stars having very long period systems (more similar to the Solar System case) do also
present an iron abundance above solar. If we take all stars having companions with periods
longer than 1000\,days and eccentricities lower than 0.3 we obtain an average $<$[Fe/H]$>$ 
of $+$0.21. A lower (but still high) value of $+$0.12 is achieved if we do not introduce 
any eccentricity limit into this sample. We caution, however, that these systems are not necessarily real Solar System analogs.

It is important to discuss the implications of this result on the planetary formation
scenarios. \citet[][]{Bos02} has shown that the formation of a giant planet as a result of 
disk instabilities is almost independent of the metallicity; this is contrary to what 
is expected from a process based in the core accretion scenario. The results presented 
here, suggesting that the probability of forming a planet (at least of the kind we are 
finding now) is strongly dependent on the metallicity of the host star, can thus be seen
as an argument for the former (traditional) core accretion scenario \citep[][]{Pol96}. 
We note that here we are talking about a probabilistic effect: the fact that the metallicity
enhances the probability of forming a planet does not mean one cannot form a planet
in a lower metallicity environment. This is mostly due to the fact that other
important and unknown parameters, like the proto-planetary disk mass and lifetime, 
do control the efficiency of planetary formation as well. Furthermore, these
results do not exclude that an overlap might exist between the two planetary
formation scenarios.

Finally, the small increase seen in the distribution of Fig.\,\ref{fig2} (right panel) 
for low metallicities is clearly not statistically significant, since only one 
planet host per bin exists in this region of the plot.

\subsubsection{Measurement precision and [Fe/H]}
\label{sec:precision}

As discussed in Paper\,I and II, the rise of the percentage of planets found as a function of
the increasing metallicity cannot be the result of any observational bias. 

In this sense, particular concern has been shown by the community regarding the fact 
that a higher metallicity will imply that the spectral lines are better defined. This 
could mean that the the final precision in radial-velocity could be better for the 
more ``metallic'' objects. However, in the CORALIE survey we always set the exposure 
times in order to have a statistical precision better than the former 
7\,m\,s$^{-1}$ instrumental long-term error\footnote{This value has only recently been improved
to about 2-3\,m\,s$^{-1}$ \citep[][]{Que01,Pep02}.}. 

A look at Fig.\,\ref{fig_errors}, where 
we plot the mean photon noise error for stars with different [Fe/H] having V magnitudes 
between 6 and 7, shows us exactly that there is no clear trend in the data. The very 
slight tendency (metal-rich stars have, in average, measurements with only about 
1-2\,m\,s$^{-1}$ better precision than metal poor stars) is definitely not able to
induce the strong tendency seen in the [Fe/H] distribution in Fig.\,\ref{fig2}, 
specially when we compare it with the usual velocity amplitude induced by the known 
planetary companions (a few tens of meters-per-second).
This also seems to be the case concerning the Lick/Keck planet search programs 
(D. Fischer, private communication).

\subsubsection{Primary mass bias}
\label{sec:mass}

The currently used planet-search surveys are based (in most cases) on samples 
chosen as volume-limited. However, the criteria to ``cut'' the sample was usually
also based on stellar temperature (i.e. $B-V$ colour). 

For a given $B-V$ (i.e. T$_{\mathrm{eff}}$),
varying the metallicity implies also changing the derived stellar mass. 
This means that we will have missed in our samples stars with very high [Fe/H] and 
low mass (they have too high $B-V$), as well as ``high'' mass objects with low [Fe/H] 
(too small $B-V$). For example, a 1.3\,M$_{\sun}$ dwarf with [Fe/H]=$-$0.4 has a 
temperature of $\sim$7000\,K \citep[][]{Sch92},
clearly outside the $B-V$ limits imposed by the CORALIE survey \citep[][]{Udr00}.
On the other side of the mass regime, a 6\,Gyr old, 0.6\,M$_{\sun}$ star with solar 
metallicity has a temperature of only $\sim$4150\,K \citep[][]{Cha99}; a star of this 
temperature not only is close to the 
border of our samples, but it is intrinsically very faint,
and thus much more difficult to follow at very high precision.

\begin{figure}[t]
\psfig{width=\hsize,file=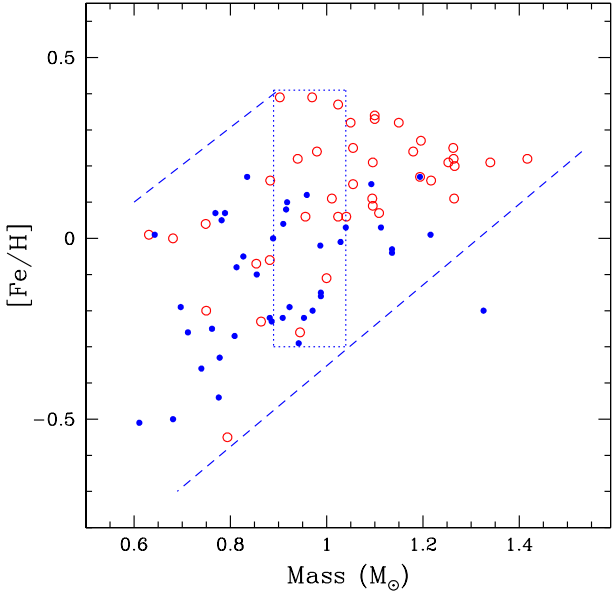}
\caption[]{Metallicity vs. stellar mass for planet (open circles) and non-planet host stars 
(filled dots) included in the CORALIE survey. As we can see from the plot, there is a strong 
bias related with the cutoff in colour of the sample. The two dashed lines represent 
simply approximate sampling limits, while the box represents a mass
region with no strong biases.}
\label{fig3b}
\end{figure}

In other words, the current samples are not really uniform in stellar masses.
That fact is well illustrated in Fig.\,\ref{fig3b}: there is clearly a trend that 
results from the definition of the sample, implying that this latter is constituted (in average) 
of more metal-poor stars as we go toward a lower mass regime. 

This strong bias makes it difficult to study the probability of planetary formation as a 
function of the stellar mass. It would be very interesting to understand e.g. if higher mass stars, that might have
slightly higher disk masses, may eventually form planets more readily\footnote{To our knowledge, there is no 
study to date of the variations in the mass of proto-planetary disks as a function of stellar mass for solar-type stars.}.
Such studies might be very important to test the planetary formation scenarios.

We can be quite sure, however, that the frequency of planetary formation for
a given stellar mass is still increasing with [Fe/H]. If we look at the plot, the region inside the box is
quite clean from the biases discussed above. And as it can be easily seen, in this region planet hosts are
still dominating the higher part of the plot. This is true even if for a given mass the higher metallicity stars
are also fainter and thus more difficult to measure. 

\subsection{Primordial source as the best explanation}

Two main different interpretations have been given to the [Fe/H] ``excess'' observed 
for stars with planets. One suggests that the high metal content is the result of the 
accretion of planets and/or planetary material into the star \citep[e.g.][]{Gon98}. 
Another simply states that the 
planetary formation mechanism is dependent on the metallicity of the proto-planetary disk: 
according to the ``traditional'' view, a gas giant planet is formed by runaway accretion 
of gas by a $\sim$10 earth-mass planetesimal. The higher the metallicity (and thus 
the number of dust particles) the faster a planetesimal can grow, and the higher the 
probability of forming a giant planet before the gas in the disk dissipates.

A third possibility is that the metallicity is in fact favoring the formation of the currently 
found (in general short period when compared to the Solar System giant planets) exoplanets \citep[][]{Gon98}. 
This could fit e.g. into the idea of planetary migration induced by the interaction of the giant planet with a
swarm of planetesimals: the higher the metallicity, the higher the number of those
minor bodies, and thus the more effective the migration could be.
However, the migration mechanisms are not very well understood, and
probably more than with the number of planetesimals or the metallicity, the migration rate 
seems to be related with the disk lifetimes, and to the (gas) disk and planetary masses \citep[][]{Tri02}. 
These variables are, however, very poorly known (do they depend e.g. on stellar mass or on the 
metallicity itself?). 

There are multiple ways of deciding between the two former scenarios (see discussion in
Paper\,II), and in particular to try to see if pollution might indeed have played
an important role in increasing the metal content of the planet host stars relative to their
non-planet host counterparts. Probably the most clear and strong argument 
is based on stellar internal structure, and in particular on the fact that material falling into 
a star's surface would induce a different increase in [Fe/H] depending on the depth of its convective envelope (where mixing can occur). 
This approach, already used by several authors \citep[][]{Lau00,San00,Gon01,Pin01,San01a,San01b,Mur02,Rei02}, has
led to somewhat opposite conclusions. 

\begin{figure}[t]
\psfig{width=\hsize,file=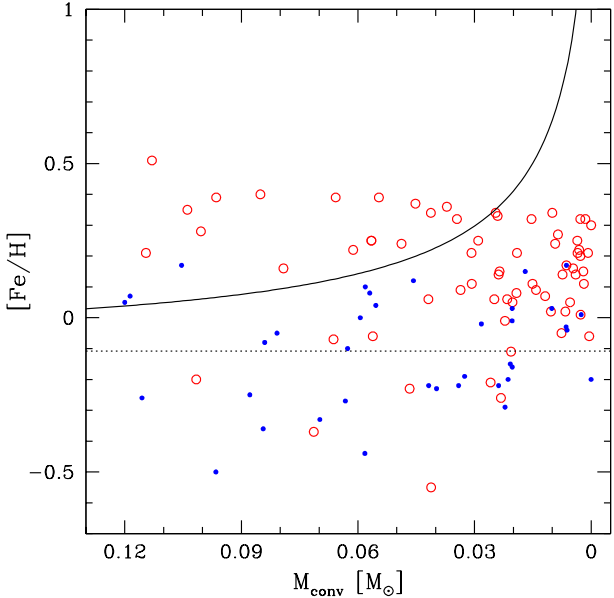}
\caption[]{Metallicity vs. convective envelope mass for stars with planets (open symbols) and field dwarfs (points). 
The [Fe/H]=constant line represents the mean [Fe/H] for the non-planet hosts stars of Fig.\,\ref{fig1}. The curved line
represents the result of adding 8 earth masses of iron to the convective envelope of 
stars having an initial metallicity equal to the non-planet hosts mean [Fe/H]. The resulting trend
has no relation with the distribution of the stars with planets.}
\label{fig3}
\end{figure}

In Fig.\,\ref{fig3} we plot the metallicity for the planet
host stars having surface gravity higher than 4.1\,dex (to avoid sub-giant stars) against their
convective envelope mass\footnote{This quantity was derived using the equations presented 
in \citet[][]{Mur01}.} (open symbols) as well as for the stars in the comparison sample
of Paper\,II (points). The dashed line represents the mean metallicity for this latter group. The curved
line represents the results of adding 8 earth masses of pure iron in the convective envelopes
of stars having an initial metallicity similar to the average metallicity of the
field star sample. 

A look at the points reveals no trend comparable to the one expected if the metallicity excess
were mainly a result of the infall of planetary material. In particular, a quick look indicates that the
upper envelope of the points is extremely constant. Furthermore, there are no stars with [Fe/H]$\ge$$+$0.5; 
this should not be the case if pollution were the main cause of the
excess metallicity. As shown by \citet[][]{Pin01}, a similar result is achieved if we replace
M$_{\mathrm{conv}}$ by T$_{\mathrm{eff}}$, since this latter quantity is well correlated with
the convective envelope mass. These authors have further shown that even non-standard models of convection and diffusion 
cannot explain the lack of a trend and thus sustain ``pollution'' as the source of the high-[Fe/H].

The analysis of Fig.\,\ref{fig3} strongly suggests that the high metal content of stars with
planets is of ``primordial'' origin. This is further supported by the fact that
the 7 planet host stars which have $\log{g}$ values lower than 4.1\,dex (probably already 
evolved stars, that have deepened their convective envelopes, diluting every metallicity 
excess that could be present at the beginning)
have a mean metallicity of 0.17\,dex, even higher than the 0.13\,dex mean value found for all the
planet hosts. This result, together with Fig.\,\ref{fig2}, implies that the metallicity 
is a key parameter controlling planet formation and evolution, and may have enormous 
implications on theoretical models (as discussed in Sect.\,\ref{sec:probability}). 

An explanation to the absence of ``important'' pollution traces 
can indeed come from arguments based on the timescales of planetary formation. Although still 
a matter of debate, near-infrared observations
suggest that circumstellar (proto-planetary) disks have lifetimes shorter than 10\,Myr \citep[e.g.][and references therein]{Hai01}. 
Considering that the disappearance of a near-IR disk (i.e. a dust disk)
also means that the gas has disappeared (a reasonable assumption), then all the processes
connected to the formation of a giant planet must happen before 10\,Myr. 
Taking the example of the Sun, after 10\,Myr its convective envelope has $\sim$0.3\,M$_{\odot}$ of
material\footnote{A 1\,M$_{\odot}$ solar metallicity star reaches the main-sequence after $\sim$30\,Myr.}. In an extreme 
case where all the solid material from the disk falls into
the star \cite[i.e. about 1\,M$_{\mathrm{Jup}}$ considering a very massive disk with 0.1\,M$_{\odot}$ of gas and dust -- ][]{Bec90}
but none H and He is accreted, the solar iron abundance would increase by only
$\sim$0.1\,dex. Even in this case, the pollution would induce a [Fe/H] variation that is 
still $\sim$0.15\,dex lower than the average difference between planet hosts and non-planet hosts.
We can thus state that after all pollution is probably not expected to make an important contribution
to the total metallicity excess\footnote{These facts can even be seen as a constraint for the timescales
of disk evolution and giant planet formation}. 
This is even stressed by the fact that higher mass stars evolve faster, and attain a
shallow convective envelope before their lower mass counterparts. This would even
strengthen the expected slope in Fig.\,\ref{fig3}: nothing is seen.

It should be noted however, that we are not excluding that, in some isolated cases,
pollution might have been able to alter more or less significantly the global metallicity of the
stars. There are some examples supporting that some planet host stars might have
suffered a limited amount of ``pollution'' \citep[e.g.][]{Gon98,Smi01,Isr01,Law01,Isr02},
although not necessarily able to change considerably the overall metal 
content \citep[see e.g.][]{Isr01,Pin01,San02,Sand02}.

It is also important to mention that here we are interested in discussing the origin of the
high metallicity of planet host stars, something that has strong implications into the theories
of planetary formation and evolution. The current results do not pretend to discuss the
general question of ``pollution'' in the solar neighborhood, a subject that
has seen some results recently published \citep[e.g.][]{Mur01,Gra01,Qui02,Gai02}. In 
particular, the discovery of possible non-planet host main-sequence binaries with different 
chemical compositions \citep[][]{Gra01} does not permit to say much regarding the planetary
``pollution'' problem.

\section{Correlations with planetary orbital parameters and minimum masses}
\label{sec:parameters}

Some hints of trends between the metallicity of the host stars and the orbital 
parameters of the planet have been discussed already in the literature. 
The usually low number of points involved in the statistics did not permit, however, to extract 
any major conclusions (see Paper\,II). Today, we dispose of about 80 high-precision and uniform metallicity 
determinations for planet host stars, a sample that enables us to look for possible 
trends in [Fe/H] with planetary mass, semi-major axis or period, and eccentricity with a higher 
degree of confidence. Let us then see what is the current situation.

\subsection{Planetary mass}
\label{sec:m2}

\begin{figure*}[t]
\psfig{width=\hsize,file=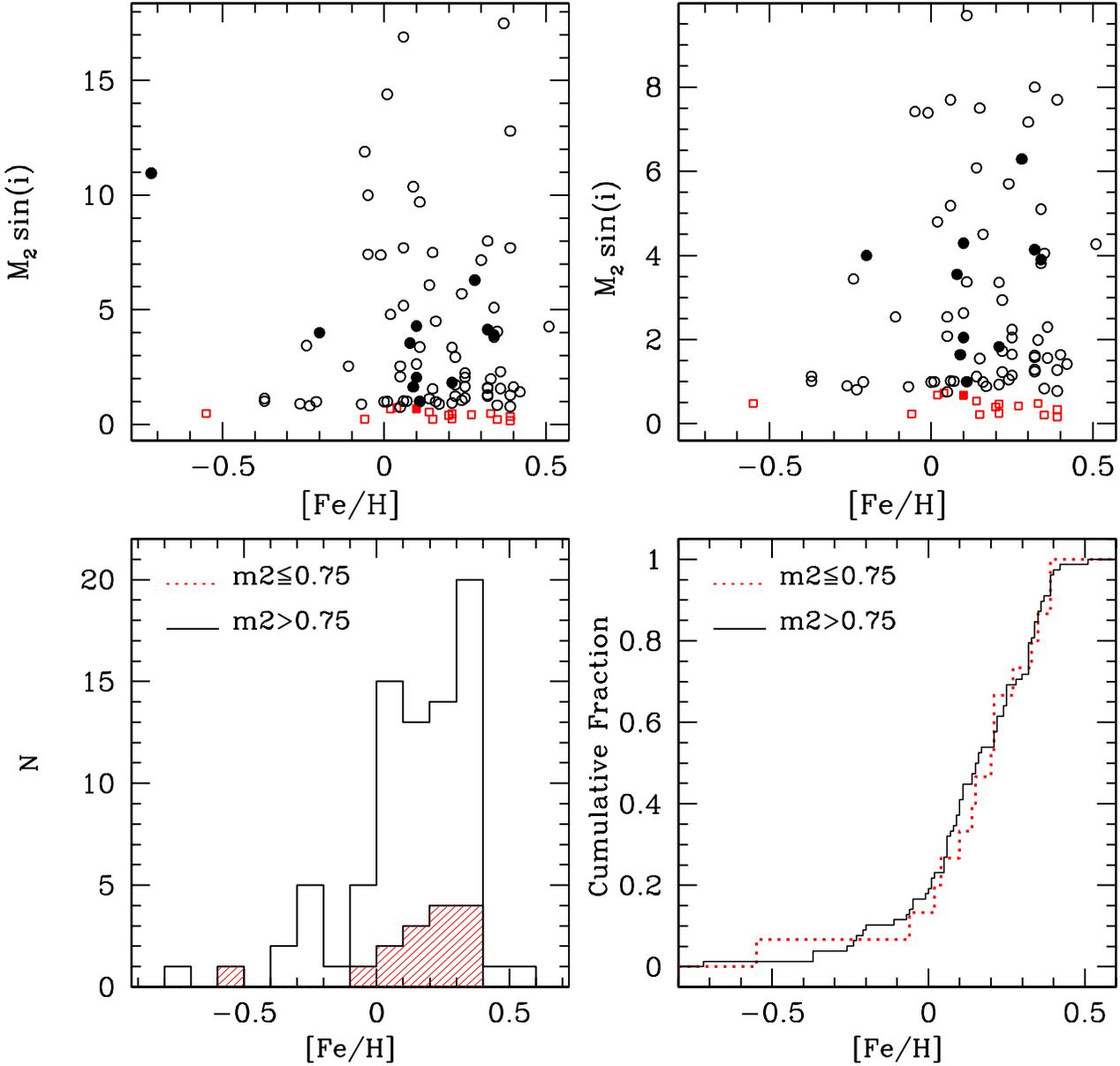}
\caption[]{{\it Upper panels}: Metallicity against minimum mass for the planetary companions
known to date whose host stars have precise spectroscopic [Fe/H] determinations. The right
plot is just a zoom of the upper plot in the region of M$_2\,\sin{i}$$<$10\,M$_{\mathrm{Jup}}$ (see text).
Different symbols go for the planets with minimum mass above (circles) 
or below (squares) 0.75\,M$_{\mathrm{Jup}}$. The filled dots represent planets in stellar systems \citep[][]{Egg02}.
{\it Lower left}: [Fe/H] distributions for stars with planets less and more massive than
0.75\,M$_{\mathrm{Jup}}$ (the hashed and open bars, respectively). {\it Lower right}: 
cumulative functions of both distributions. A Kolmogorov-Smirnov test gives a probability of 
$\sim$0.98 that both samples belong to the same distribution.}
\label{fig5}
\end{figure*}

In Fig.\,\ref{fig5} (upper left panel) we plot the minimum mass for the ``planetary'' companions 
as a function of the metallicity. A simple look at the plot gives us the impression that there 
is a lack of high mass companions to metal-poor stars. This fact, although
not clearly significant, does deserve some discussion.

If we concentrate in the region of M$_2\,\sin{i}$$<$10\,M$_{\mathrm{Jup}}$\footnote{As shown by \citet[][]{Jor01},
this is a probable upper limit for the mass of a planet} (Fig.\,\ref{fig5}, upper right panel), 
the trend  mentioned above still remains. As discussed in \citet[][]{Udr02}, this result can be seen as an evidence that to form a massive planet (at least up to a mass of $\sim$10\,M$_{\mathrm{Jup}}$) we need more metal-rich disks. For example, the upper plots of Fig.\,\ref{fig5} show
that except for HD\,114762 (a potential brown-dwarf host), all planets with masses 
above $\sim$4\,M$_{\mathrm{Jup}}$ orbit stars having metallicities similar or above to solar. 
This tendency could have to do with the time needed to build the planet seeds before 
the disk dissipates (if you form more rapidly the cores, you have more time to accrete gas around), 
or with the mass of the ``cores'' that will later on accrete gas to form
a giant planet\footnote{The higher the dust density of the disk, the bigger will be the core 
mass you might be able to form \citep[e.g.][]{Kok02}; does this mass influence the 
final mass of the giant planet?}. 

\begin{figure*}[t]
\psfig{width=\hsize,file=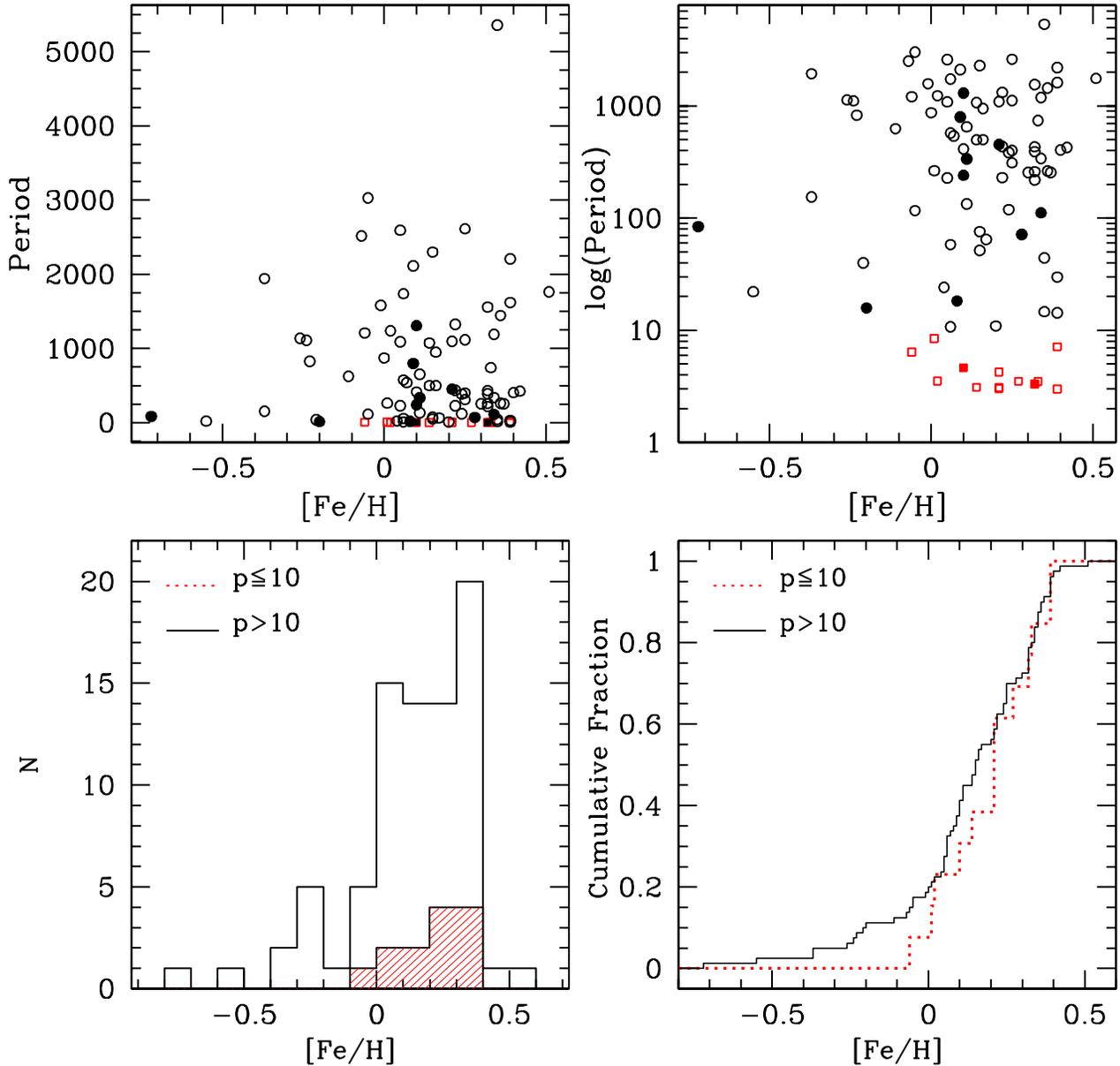}
\caption[]{{\it Upper panels}: Metallicity against orbital period for the planetary 
companions known to date and whose host stars have precise spectroscopic [Fe/H] determinations 
in linear (left) and log scales (right). Different symbols are used for
planets in orbits having periods longer and smaller than 10\,days. The filled dots represent 
planets in stellar systems \citep[][]{Egg02}.
{\it Lower left}: [Fe/H] distributions for stars with planets with orbital periods shorter and longer than
10 days (the hashed and open bars, respectively). {\it Lower right}: 
cumulative functions of both distributions. A Kolmogorov-Smirnov test gives a probability of 
$\sim$0.75 that both samples belong to the same distribution.}
\label{fig4}
\end{figure*}

In the two upper panels, the filled symbols represent planets in multiple stellar systems \citep[][]{Egg02}.
We do not see any special trend for these particular cases.

In the lower panels of Fig.\,\ref{fig5} we show the [Fe/H] distribution for the stars with low mass companions
for two different companion mass regimes. The chosen limit of 0.75\,M$_{\mathrm{Jup}}$ as a border takes
into account the striking result found by \citet[][]{Udr02b}, in the sense that planets with
masses lower than about this value have all periods shorter than $\sim$100\,days (these authors see this limit as a strong constraint for the planet migration scenarios). The Kolmogorov-Smirnov test shows that there is
no statistically significant difference between these two distributions. 
Changing the limit from 1\,M$_{\mathrm{Jup}}$ to some other value does not change the
significance of the result (nor the shape of the distributions).

\subsection{Period}
\label{sec:p}

\citet[][]{Gon98} and \citet[][]{Que00} have presented some evidences that 
stars with short-period planets (i.e. small semi-major axes) may be particularly metal-rich,
even amongst the planetary hosts. This fact could be interpreted by considering that
the migration process is able to pollute the stellar convective envelope \citep[][]{Mur98}, 
that the formation of close-in planets is favored by the metallicity, or that the 
subsequent inward orbital evolution of a newborn planet may be favored by the higher 
metallicity of the disk (e.g. by the presence of more planetesimals/planets with which the planet 
can interact). The number of planets that were known by that time was, 
however, not sufficient for us to take any definitive conclusion. In fact, in Paper\,II we
have not found that this trend was very significant, although there was still a slight 
correlation. 

In Fig.\,\ref{fig4} (upper panels) we plot the metallicity against the orbital period 
for the planets whose stars have precise spectroscopic metallicity determinations.
In the plot, the squares represent planet hosts having companions in orbits shorter 
than 10\,days (``hot jupiters''), while circles represent planets with longer orbital period.
As we can see from the plots, there seems to be a small tendency for short period planets to 
orbit more metal-rich stars (see also the two lower panels). Or, from another point of view, 
the distribution for longer period systems seems to have a low metallicity tail, not
present for the shorter period case (see the lower left panels). This tendency is clearly not significant, however 
(the Kolmogorov-Smirnov probability that both samples belong 
to the same population is of 0.75). 

\begin{figure}[t]
\psfig{width=\hsize,file=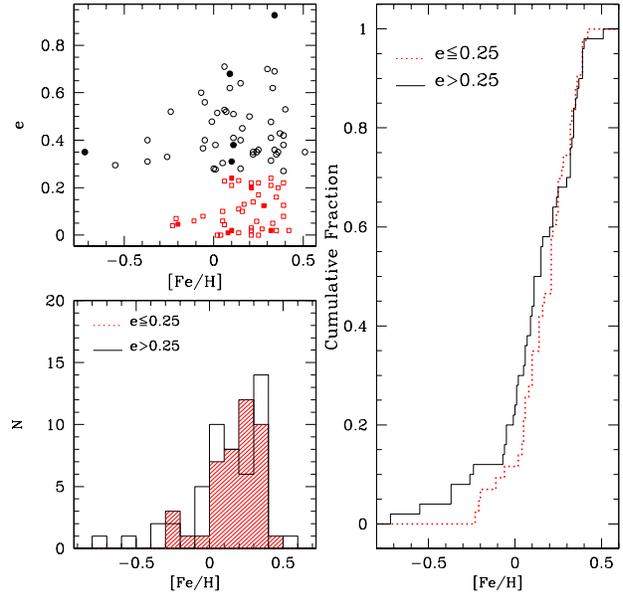}
\caption[]{{\it Upper left}: Metallicity against orbital eccentricity for the known planetary companions
whose host stars have precise spectroscopic [Fe/H] determinations. Different symbols are used for
planets in orbits having eccentricity higher or smaller than 0.25. The filled dots represent 
planets in stellar systems \citep[][]{Egg02}.
{\it Lower left}: [Fe/H] distributions for stars with planets with eccentricities lower and higher 
than 0.25 (the hashed and open bars, respectively). {\it Right}: 
cumulative functions of both distributions. A Kolmogorov-Smirnov test gives a probability of 
$\sim$0.38 that both samples belong to the same distribution.}
\label{fig6}
\end{figure}

Changing the limits does not bring further clues on any statistically
significant trend. In particular, setting the border at around 100\,days, a value that seems to have some 
physical sense (companions to stars having periods shorter than this limit have statistically lower
masses -- \citep[][]{Udr02b}) does not change the conclusions. 
We have further tried to investigate if the shape of the metallicity distributions changes if we
consider planet hosts having companions with a different range of orbital period. A very slight trend
seems to appear if we separate stars having companions with periods longer and shorter than 
1 year, in the sense that the former's [Fe/H] distribution seems to be a bit more flat. But 
at this moment this is far from being significant.

We note, however, that the two lowest metallicity stars in the samples are both in 
the $<$100\,day period regime.
These two points give us the impression (in the upper-left plot) that there
do not seem to exist any long period systems around low metallicity stars. This
impression disappears, however, if we plot the period in a logarithmic scale (upper-right panel).
No further conclusions can be taken at this moment.

As before, in the two upper panels, the filled symbols represent planets in stellar 
systems \citep[][]{Egg02}. We do not see any special trend for these particular cases.

The lack of a clear relation between orbital period and stellar metallicity might imply that
the migration mechanisms are reasonably independent of the quantity of metals in the disk.
This result might thus fit better into the scenarios
based on the migration of the planet through a gas disk \citep[e.g.][]{Gol80,Lin96} when compared to the
scenarios of migration due to interaction with a disk of planetesimals \citep[][]{Mur98}.

On the other hand, lower mass planets are supposed to migrate faster than their more massive counterparts, 
since these latter open more easily a gap in the disk, thus halting their (type-II) migration \citep[][]{Tri02}.
This is observationally supported by the discovery that there are no massive planets in short period 
orbits \citep[][]{Zuc02,Udr02}, and by the clear trend showing that planets less massive 
than $\sim$0.75\,M$_{\mathrm{Jup}}$ all follow short period trajectories \citep[][]{Udr02b}. If 
the low metallicity stars are only able to form low-mass planets (a slight trend suggested 
in the last section), metal-poor stars should have preferentially ``short''\footnote{With short 
we refer here to planets having periods that are, in average, shorter than the ones found around 
more metal-rich stars.} period planets.

\subsection{Eccentricity}
\label{sec:e}

It might also be interesting to explore whether there is any relation between the eccentricity 
of the planetary orbits and the stellar metallicity. Such an analysis is presented in Fig.\,\ref{fig6}.

As it can be seen, no special trends seem to exist. There is a slight suggestion that all the
low metallicity objects have intermediate eccentricities only. In other words, 
the more eccentric planets seem to orbit only stars with metallicity higher or comparable 
to solar, and on the opposite side of the eccentricity distribution, there seems also to 
be a lack of low eccentricity planets around metal-poor stars. 

However, as can also be seen in Fig.\,\ref{fig6} (histogram and cumulative functions), this result is
not statistically significant. We have tried to change the limits of eccentricity for the 
two [Fe/H] distributions. No further conclusions can be taken.

\section{U, V, W, and metallicity}
\label{sec:uvw}

\subsection{Kinematics of stars with planets}

There are only a few studies in the literature on the kinematics of planet host stars
\citep[][]{Gon99,Rei02,Bar02}. None of them, however, has made use of a completely
unbiased sample to compare planet and non-planet host stars.
To fill this gap, we have analyzed the spatial velocities 
distributions and velocity dispersions for the subsample of extra-solar planet host stars
that are included in the CORALIE sample, and have compared this results with space velocities 
for $\sim$1000 dwarfs that make part of the CORALIE
survey \citep[][]{Udr00} and have precise radial-velocity measurements.
We have restricted the planet sample to only those
planets belonging to the CORALIE sample in order to minimize the biases
when trying to compare planet and non-planet host stars. 

The U, V, and W velocities\footnote{U, V and W represent the usual spatial velocities 
of a star in the direction of the the galactic center, galactic rotation, and perpendicular 
to the galactic plane, respectively} were computed using CORALIE radial velocities, 
as well as coordinates and proper motions from 
Hipparcos \citep[][]{ESA97}\footnote{The values of U, V and W for the planet host stars
will be available at CDS}. 
The convention used is so that U, V and W are positive in the direction of the 
galactic center, the galactic rotation, and the north galactic pole, respectively. 
We have then corrected the velocities with respect to the Solar motion relative to the 
LSR adopting 
(U$_{\mathrm{LSR}}$,V$_{\mathrm{LSR}}$,W$_{\mathrm{LSR}}$)$_{\sun}$=(10,6,6)\,km\,s$^{-1}$ 
\citep[e.g.][]{Gon99}. 

\begin{figure}[t]
\psfig{width=\hsize,file=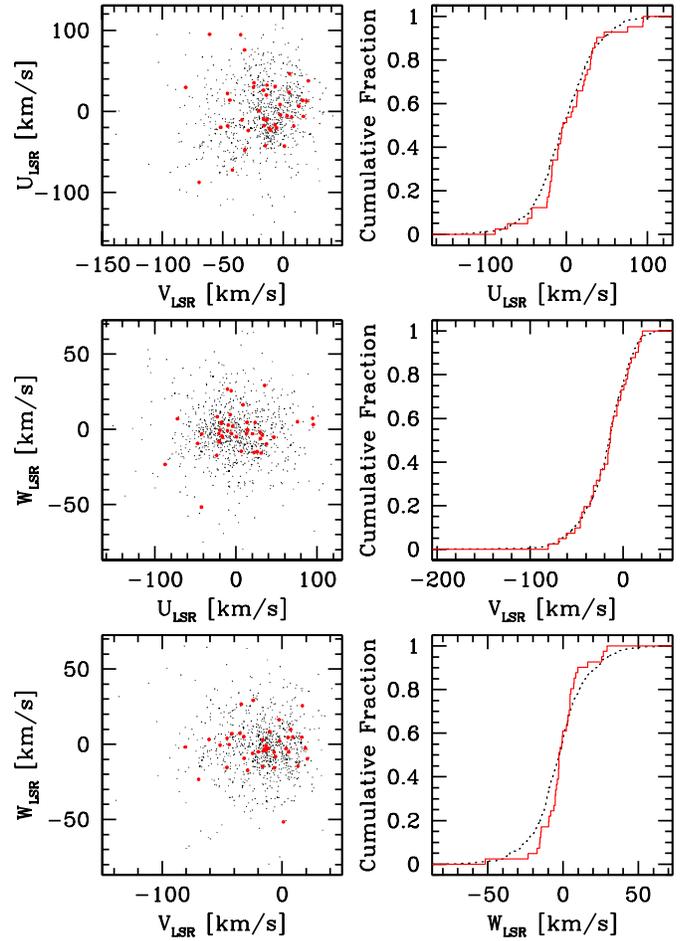}
\caption[]{{\it Left}: U$_{\mathrm{LSR}}$-V$_{\mathrm{LSR}}$, U$_{\mathrm{LSR}}$-W$_{\mathrm{LSR}}$, and 
V$_{\mathrm{LSR}}$-W$_{\mathrm{LSR}}$ diagrams for planet hosts (filled dots) and stars in the CORALIE sample (see text for
more details). {\it Right}: Cumulative functions of U$_{\mathrm{LSR}}$, V$_{\mathrm{LSR}}$, and W$_{\mathrm{LSR}}$ for the two
samples (planet hosts are the filled line, and the CORALIE sample is denoted by the
dotted line).}
\label{uvw}
\end{figure}

In Fig.\,\ref{uvw} we plot the classical U$_{\mathrm{LSR}}$-V$_{\mathrm{LSR}}$, U$_{\mathrm{LSR}}$-W$_{\mathrm{LSR}}$, and 
V$_{\mathrm{LSR}}$-W$_{\mathrm{LSR}}$ diagrams (left plots) for planet hosts (dots) and non-planet hosts (small points), as well as the
cumulative functions of U$_{\mathrm{LSR}}$, V$_{\mathrm{LSR}}$ and W$_{\mathrm{LSR}}$ (right plots) for the 
two samples. As we can see, there is no major difference between the two groups of points. This is 
supported in all cases by a Kormogorov-Smirnov test.
The only special feature to mention in this plot is that the W$_{\mathrm{LSR}}$ velocity seems to have a greater dispersion for
non-planet hosts than for planet hosts (this can be seen from the cumulative functions of W$_{\mathrm{LSR}}$ for the two samples). 

As discussed by \citet[][]{Rab98} -- see also review by \citet[][]{Gre00} -- Galactic dynamic models
imply that stars coming from the 
inner disk and influenced by the Galactic bar should present a lower dispersion in W$_{\mathrm{LSR}}$ and 
a higher U$_{\mathrm{LSR}}$. Although the former of these two trends is suggested by our data,
the higher U$_{\mathrm{LSR}}$ observed for the planet hosts (with respect to the other CORALIE sample
stars) is not significant (see also Table\,\ref{tabuvw}). 

\begin{table}[t]
\caption[]{Average space velocities and their dispersions}
\begin{tabular}{lrr}
\hline
Velocities  (km\,s$^{-1}$)    & Planet hosts      & CORALIE sample        \\
\hline
$<U_{\mathrm{LSR}}>$          &     3.2$\pm$6.0   &    $-$2.5$\pm$1.2  \\
$<V_{\mathrm{LSR}}>$          & $-$17.2$\pm$3.9   &   $-$18.3$\pm$0.8  \\
$<W_{\mathrm{LSR}}>$          &  $-$1.6$\pm$2.2   &    $-$2.6$\pm$0.6  \\
$\sigma(U_{\mathrm{LSR}})$    &    37.9$\pm$4.3   &      37.9$\pm$0.9  \\
$\sigma(V_{\mathrm{LSR}})$    &    24.5$\pm$2.8   &      25.4$\pm$0.6  \\
$\sigma(W_{\mathrm{LSR}})$    &    13.8$\pm$1.6   &      18.9$\pm$0.4  \\
\hline
\end{tabular}
\newline 
{Here we use $\sigma/\sqrt{N-1}$ for the errors in $<$U$_{\mathrm{LSR}}$$>$, $<$V$_{\mathrm{LSR}}$$>$, 
and $<$W$_{\mathrm{LSR}}$$>$, and $\sigma/\sqrt{2\,N-1}$ for the errors in $\sigma(U_{\mathrm{LSR}})$, $\sigma(V_{\mathrm{LSR}})$, and $\sigma(W_{\mathrm{LSR}})$; N equals 990 for the CORALIE sample, and 41 for the planet sample.}
\label{tabuvw}
\end{table}

In Table\,\ref{tabuvw} we list the mean U$_{\mathrm{LSR}}$, V$_{\mathrm{LSR}}$, 
and W$_{\mathrm{LSR}}$ velocities and their dispersions for the two groups of stars analyzed. 
As we can see, besides the lower dispersion in W$_{\mathrm{LSR}}$ for the planet host sample, 
the two groups do not seem to differ considerably. 
The mean total space velocity for planet and 
non-planet hosts (42.2$\pm$4.4\,km\,s$^{-1}$ and 
45.6$\pm$0.8\,km\,s$^{-1}$, respectively), and their dispersions (27.7$\pm$3.1\,km\,s$^{-1}$ and 26.4$\pm$0.6\,km\,s$^{-1}$) also do not 
show any special trend.

\begin{figure}[t]
\psfig{width=\hsize,file=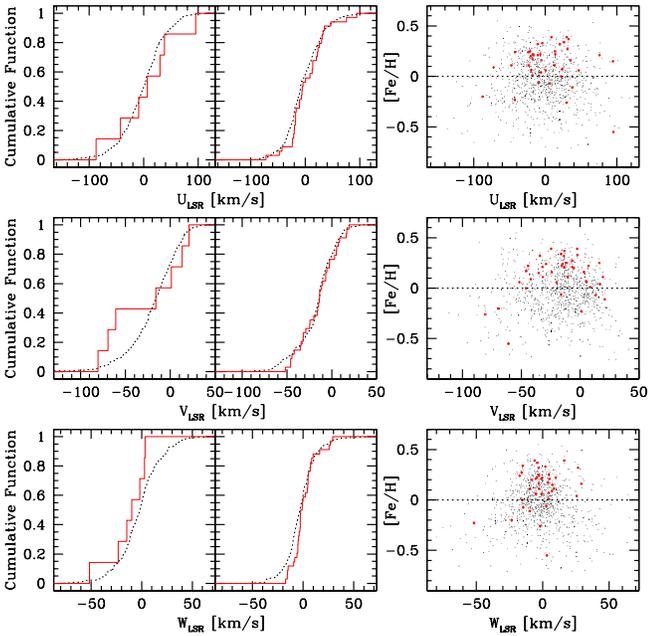}
\caption[]{{\it Right}: Metallicity as a function of the space velocities for planet and non-planet hosts. 
{\it Left}: Cumulative functions for the two samples but dividing the stars in metal rich 
([Fe/H]$>$0.0 -- right panels) and metal poor ([Fe/H]$\leq$0.0 -- left panels). Symbols as in Fig.\,\ref{uvw}.}
\label{uvwfeh}
\end{figure}

\subsection{Kinematics vs. [Fe/H]}

In Fig.\,\ref{uvwfeh} we further compare the metallicity as a function of the space velocities
for the same two samples described above. In the right panels we plot [Fe/H] as a function 
of U$_{\mathrm{LSR}}$, V$_{\mathrm{LSR}}$, and W$_{\mathrm{LSR}}$, and in the left plots we have the cumulative functions 
for U$_{\mathrm{LSR}}$, V$_{\mathrm{LSR}}$, and W$_{\mathrm{LSR}}$ as in Fig.\,\ref{uvw}, but this time separating the stars 
with [Fe/H] higher than solar (right panel) and lower than solar (left panel).
Again, no statistically significant conclusions can be drawn. Stars with planets seem to occupy 
basically the metal-rich envelope of the U$_{\mathrm{LSR}}$, V$_{\mathrm{LSR}}$, and W$_{\mathrm{LSR}}$ vs. [Fe/H] plots.

In a few words, within the statistical significance of our sample, we can say that for a given metallicity interval, 
the space velocity distribution of the planet host stars are basically the same as the one found for the 
whole planet search sample.

\section{Concluding Remarks}
\label{sec:conclusions}

In this article we present a detailed spectroscopic analysis of more than 
50 extra-solar planet host stars (to be added to the previously derived results), with the main 
goal of looking for correlations
between the stellar metallicity and the planetary orbital properties.
We have further tried to verify if planet host stars have any anomaly concerning
the space velocities. The main conclusions of the article can be summarized 
in the following way:

\begin{itemize}

\item We confirm previous results that have shown that planet host 
stars are metal-rich when compared to the average local field dwarfs. 
The addition of the new data, and the comparison of the planet host star metallicity
with the one for a large volume limited sample of field dwarfs, even
strengthens the statistical significance of the result.
Furthermore, it is shown that this result is still true if we compare
planet and non-planet host stars within a given stellar mass regime.

\item Stars with companions in the mass regime 
20\,M$_{\mathrm{Jup}}$$>$M$>$10\,M$_{\mathrm{Jup}}$ present
a wide variety of metallicities, and are up to now indistinguishable from the
stars having lower mass companions. This result
might be telling us that these ``higher'' mass companions were formed in the
same way as their lower mass counterparts. The same situation is found
for systems with more than one planet, and for planets in stellar binaries. 

\item We also confirm previous results suggesting that the probability of finding a planet 
increases with the metallicity of the star. At least 7\% of the stars in the CORALIE 
planet-search sample having [Fe/H] between 
0.3 and 0.4\,dex have a planetary companion. This frequency falls to a value of 
less then 1\% for stars with solar metallicity. We have also shown that this result cannot be
related to any observational bias.

\item Our results also strongly support the ``primordial'' source as the key parameter
controlling the high metal content of the planet host stars. ``Pollution'' does not seem
to be an ``important'' mechanism.

\item We have explored the relation between the metallicity of the host stars and the
orbital parameters of the planets. We have found some indications (not statistically significant) that metal-poor 
stars might be mainly able to form low mass planets, a result previously discussed by \citet[][]{Udr02}. 
In an even less significant sense, the low [Fe/H] stars seem to be orbited (in average) by planets 
having intermediate eccentricities. As for the orbital period, very short period planets also 
seem to orbit preferentially metal-rich stars,
but the difference with respect to their longer period counterparts is clearly not significant.
In any case, the results seem to be telling us that the metallicity is not playing a very important
role in determining the final orbital characteristics of the discovered exoplanets. The metal content of the disk does not seem to be a crucial parameter controlling the migration processes. 
But it will be very interesting to follow the results as more low mass and long period planets,
more similar to the Solar System giants, become common in the planet discovery lists.

\item No significant differences are found when comparing planet hosts and non-planet hosts
in plots relating U, V, W and [Fe/H]. In general, planet host stars seem to occupy
the metal-rich envelope of the plots.

\end{itemize}

The results presented above seem to support a scenario where the formation of giant planets,
or at least of the type we are finding now, is particularly dependent on the [Fe/H] of
the primordial cloud. In other words, and as already discussed in Paper II, the metallicity seems 
to play a crucial role in the formation of giant planets, as it is indicated by the shape of the
metallicity distribution. 

Our results further support the core accretion scenario \citep[e.g.][]{Pol96} against the disk instability 
model \citep[][]{Bos00} as the ``main'' mechanism of giant-planetary formation. 
In fact, \citet[][]{Bos02} has shown that contrarily to the core-accretion models, the efficiency of
planetary formation in the case of the disk instability should not strongly depend on the 
metallicity of the disk. In other words, if that were the case, we should probably not see
an increase in the frequency of planets as a function of the metallicity: such a
trend is clearly seen in our data. We note, however, that the current data does not discard that both 
situations can occur.

In the future, it is important to try to explore and compare the abundances of other metals (besides light elements)
in planet and non-planet host stars. 
In this context, there are already a few results published \citep[][]{San00,Smi01,Gon01,Tak01,Sad02}, but up to now
there have been no published uniform studies, a matter that clearly makes a comparison difficult.
We are currently working on this problem, in order to offer a uniform comparison
between planet and non-planet hosts for other elements in a similar way as was done here for [Fe/H].

As seen in the current work, the increase in the number of known exoplanets
is permitting the development of different sorts of statistical studies.
However, in face of the number of variables that might be interrelated 
(e.g. metallicity, and the various orbital parameters) it is still
difficult to take any statistically significant conclusions. 
A clear improvement of the current situation can only be
achieved if the kind of studies presented here are continued as new
planets are found.

\begin{acknowledgements}
  We would like to thank Nami Mowlavi for the important help in determining the stellar masses,
  and C. Perrier for obtaining ELODIE spectra for two of the stars.
  We wish to thank the Swiss National Science Foundation (Swiss NSF) for the continuous support 
  to this project. Support from Funda\c{c}\~ao para a Ci\^encia e Tecnologia 
  (Portugal) to N.C.S. in the form of a scholarship is gratefully acknowledged.
\end{acknowledgements}

\end{document}